\documentclass[aps,pra,twocolumn,superscriptaddress,showpacs,amsmath,amssymb]{revtex4}

\usepackage{graphicx}
\usepackage{bm}
\usepackage{amssymb}

\begin{document}

\title{Angular Momenta and Spin-Orbit Interaction of Nonparaxial Light in Free Space}

\author{Konstantin Y. Bliokh}
\email{k.bliokh@gmail.com}
\affiliation{Applied Optics Group, School of Physics, National University of Ireland, Galway, Galway, Ireland}

\author{Miguel A. Alonso}
\affiliation{The Institute of Optics, University of Rochester, Rochester, New York, 14627, USA}
\affiliation{Department of Applied Physics, Aalto University, PO Box 13500, FI-00076 Aalto, Finland}

\author{Elena A. Ostrovskaya}
\affiliation{ARC Centre of Excellence for Quantum-Atom Optics and
Nonlinear Physics Centre, Research School of Physics and Engineering, The Australian National University,
Canberra ACT 0200, Australia}

\author{Andrea Aiello}
\affiliation{Max Planck Institute for the Science of Light, G$\ddot{u}$nter-Scharowsky-Stra$\beta$e 1/Bau 24, 91058 Erlangen, Germany}
\affiliation{Institute for Optics, Information and Photonics, University Erlangen-N$\ddot{u}$rnberg, Staudtstr. 7/B2, 91058 Erlangen, Germany}


\begin{abstract}
We give an exact self-consistent operator description of the spin and orbital angular momenta, position, and spin-orbit interactions of nonparaxial light in free space. Both quantum-operator formalism and classical energy-flow approach are presented. We apply the general theory to symmetric and asymmetric Bessel beams exhibiting spin- and orbital-dependent intensity profiles. The exact wave solutions are clearly interpreted in terms of the Berry phases, quantization of caustics, and Hall effects of light, which can be readily observed experimentally.
\end{abstract}

\pacs{42.50.Tx, 03.65.Vf, 42.25.Ja, 42.15-i}

\maketitle

\section{Introduction}

The problem of the identification of the spin and orbital parts of the angular momentum (AM) of an electromagnetic wave has a long history and has posed fundamental difficulties in both quantum electrodynamics and classical optics \cite{AB,EN,BA,B}. 

It is known that the photon AM operator in the momentum (plane-wave) representation has the form \cite{AB}:
\begin{equation}\label{eqn:1}
\mathbf{\hat J} =  - i\left( {\mathbf{k} \times \partial _\mathbf{k} } \right) + \mathbf{\hat S} \equiv \mathbf{\hat L} + \mathbf{\hat S}~.
\end{equation}
Here the orbital part is $\mathbf{\hat L}=\mathbf{\hat r} \times \mathbf{\hat p}$ ($\mathbf{\hat p}={\bf k}$, $\mathbf{\hat r}=i{\partial_{\bf k}}$, ${\bf k}$ is the wave vector, and we use units $\hbar=c=1$), whereas $\mathbf{\hat S}$ is the spin-1 operator given by $3\times 3$ matrices $(\hat{S}_a)_{ij}=-i\epsilon_{aij}$ ($\epsilon_{aij}$ is the Levi-Civita symbol) that act on the Cartesian components of the wave electric field. Canonical orbital AM (OAM) and spin AM (SAM) operators, $\mathbf{\hat L}$ and $\mathbf{\hat S}$, satisfy $so(3)$ algebra and generate rotations in spatial and polarization degrees of freedom, respectively. However, ``\textit{the separation of the total AM into orbital and spin parts has restricted physical meaning. ... States with definite values of OAM and SAM do not satisfy the condition of transversality in the general case.}'' \cite{AB}. In 1994, Van Enk and Nienhuis put forward an alternative, non-canonical AM separation, where the modified spin and orbital parts are measurable and consistent with the transversality of the wave, although they are not generators of rotations \cite{EN}.

In classical optics, the two parts of Eq.~(\ref{eqn:1}) can be unambiguously associated with the OAM and SAM for \textit{paraxial} light, where the eigenmodes of $\hat{L}_z=-i\partial_{\phi}$ ($\phi$ is the azimuthal angle in ${\bf k}$ space) and $\hat{S}_z$ are circularly polarized vortex beams with the corresponding quantum numbers $\ell = 0,\pm 1,\pm 2,...$ (topological charge of the vortex $e^{i\ell\phi}$) and $\sigma=\pm 1$ (helicity) \cite{OAM}. 
However, for \textit{non-paraxial} fields the identification of OAM and SAM meets serious difficulties \cite{EN,BA,B}. Calculations based on the recently suggested division of the Poynting energy flow into spin and orbital parts \cite{BA,Berry2009,Li,Beksh} show that the non-paraxial correction to the OAM is proportional to $\sigma$ rather than to $\ell$ \cite{BA,Li}. This resulted in the conclusion that ``\textit{in the general non-paraxial case there is no simple separation into $\ell$-dependent orbital and $\sigma$-dependent spin component of AM}'' \cite{BA}.

In this paper we re-examine the problem and give an exact self-consistent solution in terms of both the fundamental photon operators and classical energy flows. The identification of the well-defined measurable OAM and SAM of light is shown to be closely related to the analogous problem for the \textit{position} of localized photons \cite{Pryce,Photon,Hawton}. Our approach generalizes and unifies previously disjointed results: (i) non-canonical OAM and SAM operators obtained earlier for the second-quantized fields \cite{EN}; (ii) non-commutative photon position operator and Berry monopole field in momentum space \cite{Pryce,Photon}; and (iii) separation of the spin and orbital parts of Poynting energy flows \cite{Berry2009,Li,Beksh}. We find that the $\sigma$-dependent non-paraxial part of the OAM arises from Berry-phase terms describing the \textit{spin-orbit interaction} (SOI) of light. A similar effect occurs dynamically upon \textit{spin-to-orbital AM conversion} in focusing and scattering of polarized light \cite{Beksh,AMC1,AMC2,Oscar}. Other manifestations of the SOI are the \textit{spin} \cite{SHE1,SHE2,Half,Aiello} and \textit{orbital} \cite{OHE1,OHE2,OHE3,OHE4}  \textit{Hall effects of light} (i.e., $\ell$- and $\sigma$-dependent transverse shifts of the field center of gravity) that are described by our position operator and take place even in free space \cite{Aiello,OHE4}. We apply the general theory to vector Bessel beams, for which the fundamental operators manifest themselves in immediately observable $\ell$- and $\sigma$-dependent intensity distributions. The exact wave results are also explained in terms of the underlying geometrical-optics rays and caustics.

\section{Operator formalism}

We consider an electromagnetic field in free space, characterized by its plane-wave electric-field spectrum ${\bf \tilde E}\left( {\bf k} \right)$ without evanescent modes. 
The SOI of light originates from the transversality constraint, ${\bf k} \cdot {\bf \tilde E} = 0$, which couples polarization to the wave vector and reduces the full 3D vector space of the electric field components to the 2D subspace of the components tangential to a sphere of directions in ${\bf k}$-space. The operators ${\bf \hat L}$ and ${\bf \hat S}$ do not keep this subspace invariant, i.e., their action on a transverse mode results in a non-zero longitudinal component \cite{AB,EN}. However, this subspace is invariant for the total AM operator ${\bf \hat J}$, and one can divide it into two parts consistent with the transversality condition: 
\begin{equation}\label{eqn:2}
{\bf \hat J} = {\bf \hat L'} + {\bf \hat S'},~{\bf \hat L'} = {\bf \hat L} - {\bm \kappa } \times \left( {{\bm \kappa } \times {\bf \hat S}} \right),~{\bf \hat S'} = {\bm \kappa }\left( {{\bm \kappa } \cdot {\bf \hat S}} \right),
\end{equation}
where ${\bm \kappa } = {\bf k}/k$ and the modified OAM and SAM operators ${\bf \hat L'}$ and ${\bf \hat S'}$ 
can be regarded as projections of the operators ${\bf \hat L}$ and ${\bf \hat S}$ onto the transversality subspace \cite{EN}.

The modified SAM operator ${\bf \hat S'}$ is proportional to the helicity operator $\hat \sigma  = {\bm \kappa } \cdot {\bf \hat S}$, whereas the OAM operator can be written as ${\bf \hat L'} = {\bf \hat r'} \times {\bf k}$ with
\begin{equation}\label{eqn:3}
{\bf \hat r'} = {\bf \hat r} + \frac{{\bf k} \times {\bf \hat S}}{k^2 } = i\partial _{\bf k}  + \frac{{\bf k} \times {\bf \hat S}}{k^2 }~.
\end{equation}
The modified position operator (\ref{eqn:3}) has been considered in the context of photon localization and Berry phase \cite{Pryce,Photon,Hawton}. It describes the observable center of gravity of the field and brings about the space non-commutativity with the monopole term in ${\bf k}$-space:
\begin{equation}\label{eqn:4}
\left[ {\hat r'_i ,\hat r'_j } \right] =  - i\epsilon _{ijl} \hat \sigma \frac{k_l }{k^3 }~.
\end{equation}
The operators ${\bf \hat L'}$ and ${\bf \hat S'}$ do not satisfy the $so(3)$ AM algebra and have unusual commutation relations:
\begin{equation}\label{eqn:5}
\left[ {\hat S'_i ,\hat S'_j } \right] = 0,~\left[ {\hat L'_i ,\hat L'_j } \right] = i\epsilon _{ijl} (\hat L'_l -\hat S'_l),~
\left[ {\hat L'_i ,\hat S'_j } \right] = i\epsilon _{ijl} \hat S'_l.
\end{equation}
At the same time, the modified operators transform as vectors under rotations: $\left[ {\hat J_i ,\hat O'_j } \right] = i\epsilon _{ijl} \hat O'_l$, ${\bf \hat O'} = {\bf \hat L'}$, ${\bf \hat S'}$, and ${\bf \hat r'}$. The commutation relations (\ref{eqn:5}) unveil the similarity of operators ${\bf \hat L'}$ and ${\bf \hat S'}$ to those obtained for the second-quantized fields in \cite{EN}. Although they do not generate rotations, it is suggested that they do correspond to observable continuous values of the OAM and SAM of a non-paraxial transverse field \cite{EN}.

Remarkably, in the helicity representation the matrix components of the operators (\ref{eqn:2}) and (\ref{eqn:3}) become diagonal. We introduce spherical coordinates $\left(\theta,\phi,k\right)$ with basic vectors $\left( {{\bf e}_\theta  ,{\bf e}_\phi  ,{\bm \kappa }} \right)$ in ${\bf k}$-space, so that the free electric field has only $\left( {{\bf e}_\theta  ,{\bf e}_\phi} \right)$-components. The helicity basis of circular polarizations corresponds to the basic vectors ${\bf e}^ \pm   = e^{\pm im\phi } \left({{\bf e}_{\theta} \pm i{\bf e}_{\phi} }\right)/{\sqrt 2 }$, where $e^{\pm im\phi }$ is an arbitrary gauge factor \cite{Hawton}. Transition from the global Cartesian field components $( {\tilde E_x ,\tilde E_y ,\tilde E_z } )^T$ to the helicity amplitudes $( {\tilde E^ +  ,\tilde E^ -  ,\tilde E_\parallel  } )^T$ is realized via the local unitary transformation $\hat U\left( {\theta ,\phi } \right) = \hat R_z \left( { - \phi } \right)\hat R_y \left( { - \theta } \right)\hat R_z \left( {m\phi } \right)\hat V $, where $\hat R_a \left( \alpha  \right) = e^{i\alpha \hat S_a }$ is the matrix of rotation by an angle $\alpha$ with respect to the $a$-axis, whereas $\hat V$ is the constant transformation from linear- to circular-polarization basis. Making the transformation of operators (\ref{eqn:2}) and (\ref{eqn:3}) to the helicity basis, ${\bf \hat O'} \to \hat U^\dag  {\bf \hat O'}\hat U$, we obtain:
\begin{eqnarray}
\label{eqn:6}
{\bf \hat S'} = {\bm \kappa }\hat \sigma~,~~{\bf \hat L'} = 
- i{\bf k} \times \partial _{\bf k}  - {\bf \hat A}_B  \times {\bf k}~,\\
\label{eqn:7}
{\bf \hat r'} = i\partial _{\bf k}  - {\bf \hat A}_B~,~~{\bf \hat p} = {\bf k}~,~~{\hat w}=\omega
\end{eqnarray}
Here we included the momentum and energy operators, ${\bf \hat p}$ and ${\hat w}$ (which are unaffected by the transformations), $\omega$ is the frequency, the helicity is diagonal: $\hat \sigma  = {\mathop{\rm diag}\nolimits} \left( {1, - 1,0} \right)$, and 
\begin{equation}\label{eqn:8}
{\bf \hat A}_B  =  - \frac{{{\bf k} \times {\bf \hat S}}}{{k^2 }} - i\hat U^\dag  \partial _{\bf k} \hat U = \frac{{m - \cos \theta }}{{k\sin \theta }}\hat \sigma {\bf e}_\phi
\end{equation}
is the Berry gauge field (connection) which corresponds to the monopole curvature ${\bf \hat F}_B  = \partial _{\bf k}  \times {\bf \hat A}_B  = \hat \sigma \,{\bf k}/k^3$ \cite{Photon,Hawton}. Hereafter we choose the gauge $m=1$, which corresponds to the absence of the phase singularity (Dirac string) along the positive $z$-axis in Eq.~(\ref{eqn:8}) \cite{Hawton}, allowing a smooth transition to the paraxial case, $\theta  \to 0$.

It is worth noticing that the transformation to the helicity basis is associated with the transition to the local coordinate frame with the $z$-axis attached to the current ${\bf k}$-vector, which induces pure gauge Coriolis-type potential ${\bf \hat A}= - i\hat U^\dag  \partial _{\bf k} \hat U$, i.e., $( {{\bf \hat A} } )_{ij}  =  - i{\bf e}_i^*  \cdot \left( {\partial _{\bf k} } \right){\bf e}_j $, where $\mathbf{e}_{1,2,3}  \equiv \left( {\mathbf{e}^ +  ,\mathbf{e}^ -  ,\bm{\kappa }} \right)$ \cite{Photon,Coriolis}. At the same time, non-canonical operators and commutation relations (\ref{eqn:2})--(\ref{eqn:5}) essentially owe their origin to the projection onto the transversality subspace, which is equivalent to the diagonalization of the potential ${\bf \hat A}$ \cite{Photon,Coriolis}: ${\bf \hat A}_B={\rm dg}{\bf \hat A}$, i.e.,
\begin{equation}\label{eqn:9}
( {{\bf \hat A}_B } )_{ij}  =  - i{\bf e}_i^*  \cdot \left( {\partial _{\bf k} } \right){\bf e}_j \delta _{ij}~.
\end{equation}
While such diagonalization (which uncouples the two helicity components) is an adiabatic approximation for a nearly-transverse paraxial wave beam propagating in an inhomogeneous medium \cite{Coriolis, Berry1987}, it is exact for transverse plane waves in free space where the helicities are truly independent.

The measurable expectation (mean) values of the OAM, SAM, coordinate, momentum, and energy obtained from the diagonal operators (\ref{eqn:6})--(\ref{eqn:8}) can be written as
\begin{eqnarray}
\label{eqn:10}
 {\bf S} = 
\left\langle {\tilde E^\sigma  } \right| \sigma {\bm \kappa }\left| {\tilde E^\sigma  } \right\rangle,~\\
\label{eqn:11}
{\bf L} = 
\left\langle {\tilde E^\sigma  } \right|{\bf \hat L}\left| {\tilde E^\sigma  } \right\rangle  - \left\langle {\tilde E^\sigma  } \right| \sigma {\bf A}_B \times {\bf k} \left| {\tilde E^\sigma  } \right\rangle,\\
\label{eqn:12}
{\bf R} = 
\left\langle {\tilde E^\sigma  } \right|i\partial _{\bf k} \left| {\tilde E^\sigma  } \right\rangle  - \left\langle {\tilde E^\sigma  } \right| \sigma {\bf A}_B  \left| {\tilde E^\sigma  } \right\rangle,\\
\label{eqn:13}
{\bf P}= \left\langle {\tilde E^\sigma  } \right|{\bf k}\left| {\tilde E^\sigma  } \right\rangle,~
W= \left\langle {\tilde E^\sigma  } \right|\omega\left| {\tilde E^\sigma  } \right\rangle.
\end{eqnarray}
Here ${\bf A}_B  = {\bf e}_\phi k^{ - 1} \left( {1 - \cos \theta } \right)/\sin \theta$, convolution implies summation over $\sigma=\pm 1$ and integration in the ${\bf k}$-space, and we assume normalization $N=\left\langle {\tilde E^\sigma  } \right|\left. {\tilde E^\sigma  } \right\rangle=1$ (see Appendix for details). While the SAM is purely \textit{intrinsic} (origin-independent), the OAM, in general, has both intrinsic and \textit{extrinsic} contributions \cite{Extrinsic}:
\begin{equation}\label{eqn:14}
{\bf L}^{\rm ext}={\bf R}\times {\bf P}~,~~{\bf L}^{\rm int}={\bf L} - {\bf L}^{\rm ext}~.
\end{equation}

Equations (\ref{eqn:10})--(\ref{eqn:14}) contain all the main observable results related to the AM and SOI of light. First, the $\sigma$-dependent non-paraxial Berry-phase term in ${\bf L}$ should be associated with the \textit{spin-to-orbit AM conversion} \cite{Li,Beksh,AMC1,AMC2,Oscar}. Particular cases of this term have appeared in \cite{BA,Li,Beksh}. Second, the \textit{orbital} \cite{OHE1,OHE2,OHE3,OHE4} and \textit{spin} \cite{SHE1,SHE2,Half,Aiello} \textit{Hall effects of light} are described by the two terms in the position of the center of gravity, Eq.~(\ref{eqn:12}). Indeed, for a symmetric vortex beam propagating along the $z$-axis, the transverse coordinates of the center of gravity vanish, $(X,Y)=0$, 
after integration over $\phi$, but any asymmetry of the field distribution along, say, the $x$-axis immediately causes an $\ell$- and $\sigma$-dependent shift along the orthogonal $y$-axis, $Y\neq 0$ together with tilt $P_x\neq 0$ (see the example in Section IV).

We emphasize that our results (\ref{eqn:6})--(\ref{eqn:14}) are \textit{exact} and no approximations were made. They are equivalent to application of the canonical operators ${\bf \hat L}$, ${\bf \hat S}$, and ${\bf \hat r}$ to the laboratory-frame field components $( {\tilde E_x ,\tilde E_y ,\tilde E_z } )^T$ supplied with the transversality condition.

\section{Energy flow approach}

Remarkably, the same results, Eq.~(\ref{eqn:10})--(\ref{eqn:14}), can be derived from an approach based on the separation of the spin and orbital parts in the Poynting energy flow \cite{Berry2009,Li}. Let us consider a monochromatic beam-like field propagating in the positive $z$-direction. Pecularities of the $(2+1)D$ formalism for such a problem are discussed in the Appendix, Eqs.~(\ref{eqn:A6})--(\ref{eqn:A9}). 

The transverse center of gravity (\ref{eqn:A9}) obtained in the momentum representation from operator equation (\ref{eqn:12}) can be equally derived from the traditional coordinate-representation definition 
\begin{equation}\label{eqn:31}
\mathbf{R}_ \bot  \left( z \right) = \frac{1}{g^2} \int {\mathbf{r}_ \bot  \left| {\mathbf{E}\left( {\mathbf{r}_ \bot  ,z} \right)} \right|^2 \,} d^2 \mathbf{r}_ \bot~,
\end{equation}
where ${\bf r}_{\bot}=(x,y)$, $g=\sqrt{2\omega/\varepsilon_0}$, and we used normalization $\int {  \left| {\mathbf{E}\left( {\mathbf{r}_ \bot  ,z} \right)} \right|^2 \,} d^2 \mathbf{r}_ \bot = g^2$ corresponding to $N=1$ (see Appendix). Substituting here Fourier representation (\ref{eqn:A6}) with the helicity-basis expansion (\ref{eqn:A2}), and using $\int {\mathbf{r}_ \bot  e^{i\mathbf{k}_ \bot   \cdot \mathbf{r}_ \bot  } d^2 \mathbf{r}_ \bot  }  = \left( {2\pi } \right)^2 \delta ^2 \left( {\mathbf{k}_ \bot  } \right)\partial _{\mathbf{k}_ \bot  }$ (${\bf k}_\bot =(k_x , k_y )$) together with expression (\ref{eqn:9}) for the Berry connection, we arrive at Eq.~(\ref{eqn:A9}) which is equivalent to Eq.~(\ref{eqn:12}).

To derive the linear and angular momenta of the field, we use the Poynting vector which determines the momentum density (energy flow) \cite{Born}: 
\begin{equation}\label{eqn:32}
{\bm{\pi}} = \frac{1}{g^2} \operatorname{Im} \left[ {{{\mathbf{E}}^*} \times \left( {\nabla  \times {\mathbf{E}}} \right)} \right]~.
\end{equation}
Substituting here the Fourier decomposition (\ref{eqn:A6}), we obtain
\begin{equation}\label{eqn:33}
\bm{\pi} = {\mathop{\rm Re}\nolimits} \iint {e^{i\left( {{\bf k} - {\bf k'}} \right) \cdot {\bf r}} \tilde E^{\sigma\prime *}  \tilde E^\sigma  {\bf e}^{\sigma\prime *} \times \left( {{\bf k} \times {\bf e}^\sigma  } \right)} \frac{d^2 {\bf k}^{\prime}_\bot}{2\pi} {\frac{d^2 {\bf k}_\bot}{2\pi} },
\end{equation}
where $\tilde E^{\sigma\prime}   = \tilde E^\sigma  \left( {{\bf k'}} \right)$, ${\bf e}^{\sigma\prime}   = {\bf e}^\sigma  \left( {{\bf k'}} \right)$, and summation over $\sigma =\pm 1$ is implied hereinafter. After some calculations the total momentum density (\ref{eqn:33}) can be decomposed into the orbital and spin parts as suggested in \cite{Berry2009,Li}, $\bm{\pi}=\bm{\pi}^{\rm o}+\bm{\pi}^{\rm s}$:
\begin{equation}\label{eqn:34}
{\bm \pi}^{\rm o} = \iint {e^{i {\bf K}^{-} \cdot {\bf r}} {\tilde E}^{\sigma} {\tilde E}^{\sigma\prime *} \left( {\bf e}^{\sigma} \cdot {\bf e}^{\sigma\prime *} \right) \frac{{\bf K}^+}{2}} 
\frac{{d^2 {\mathbf{k}}^{\prime}_\bot }}
{{2\pi }} \frac{{d^2{\mathbf{k}}_\bot}}
{{2\pi }},
\end{equation}
\begin{equation}\label{eqn:35}
{\bm \pi}^{\rm s} = \iint {
e^{i {\bf K}^{-} \cdot {\bf r}}{\tilde E}^{\sigma} {\tilde E}^{\sigma\prime *} \left( {\bf e}^{\sigma} \times {\bf e}^{\sigma\prime *} \right)   \times \frac{{\bf K}^-}{2}} \frac{d^2 {\bf k}^{\prime}_\bot }
{2\pi } \frac{d^2 {\bf k}_\bot }{2\pi },
\end{equation}
where ${\bf K}^{\pm}={\mathbf{k}} \pm {\mathbf{k'}}$. Note that in the decomposition (\ref{eqn:33})--(\ref{eqn:35}) the two helicity components are exactly separated without an interference term \cite{Berry2009}. Also, the ``electro-magnetic democracy'' discussed by Berry \cite{Berry2009} is accomodated, because switching to magnetic-field plane-wave helicity amplitudes, ${{\tilde E}^\sigma } \rightarrow {{\tilde H}^\sigma } =  -i\sigma {{\tilde E}^\sigma }$, keeps Eqs.~(\ref{eqn:33})--(\ref{eqn:35}) invariant.
 
The linear momentum of the beam (per unit $z$-length) is given by the 2D space integration of the momentum densities (\ref{eqn:33})--(\ref{eqn:35}): $\mathbf{P} = \int {\bm{\pi }\,d^2 \mathbf{r}_ \bot  }$. In doing so, we find that the spin momentum density makes no contribution to the linear momentum \cite{Li}: ${\bf P}^{\rm s} = \int {\bm{\pi }^{\rm s}  \,d^2 \mathbf{r}_ \bot = 0}$, while the orbital contribution yields
\begin{equation}\label{eqn:36}
\mathbf{P}^{\rm o}   = \mathbf{P} = \int {\bm{\pi }^{\rm o}  d^2 \mathbf{r}_ \bot   = \int {\mathbf{k}\left| {\tilde E^\sigma  } \right|^2 } d^2 \mathbf{k}_ \bot  }~,
\end{equation}
which obviously coincides with Eq.~(\ref{eqn:13}). 

The SAM and OAM of the beam (per unit $z$-length) can be obtained by the 2D space integration of their densities, i.e.: 
\begin{equation}\label{eqn:37}
\mathbf{L} = \int {\mathbf{r} \times \bm{\pi }^{\rm o}  d^2 \mathbf{r}_ \bot  }~,~~ 
\mathbf{S} = \int {\mathbf{r} \times \bm{\pi }^{\rm s}  d^2 \mathbf{r}_ \bot  }~.
\end{equation}
Substituting Eqs.~(\ref{eqn:34}) and (\ref{eqn:35}) into Eqs.~(\ref{eqn:37}) and employing properties of Fourier integrals, we arrive at
\begin{equation}\label{eqn:38}
\mathbf{S} = i\int { ({\bf e}^{\sigma}\times {\bf e}^{\sigma *}) \left| {\tilde E^\sigma  } \right|^2 d^2 \mathbf{k}_ \bot  }
= \int {\sigma \bm{\kappa }\left| {\tilde E^\sigma  } \right|^2 d^2 \mathbf{k}_ \bot  }~,
\end{equation}
\begin{widetext}
\begin{equation}\label{eqn:39}
\mathbf{L} = \int {\tilde E^{\sigma *} \mathbf{e}^{\sigma *}  \cdot \left( { - i\mathbf{k} \times \partial _\mathbf{k} } \right)\tilde E^\sigma  \mathbf{e}^\sigma  d^2 \mathbf{k}_ \bot  }  
   =  \int {\tilde E^{\sigma *} \left( -i{\mathbf{k} \times \partial _\mathbf{k} } \right)\tilde E^\sigma  d^2 \mathbf{k}_{\bot}  }  - \int { \sigma (\mathbf{A}_B  \times \mathbf{k}) \left| {\tilde E^\sigma  } \right|^2  d^2 \mathbf{k}_{\bot} },  
\end{equation}
\end{widetext}
where identity ${\bf e}^{\sigma *}\times {\bf e}^{\sigma}=i\sigma{\bm \kappa}$ and Eq.~(\ref{eqn:9}) were used.
Clearly, the values of SAM and OAM, Eqs.~(\ref{eqn:38}) and (\ref{eqn:39}), derived from the Poynting energy flows are in perfect agreement with our operator formalism, Eqs.~(\ref{eqn:10}) and (\ref{eqn:11}) (see also Appendix). 

\begin{figure}[t]
\includegraphics[width=8.6cm, keepaspectratio]{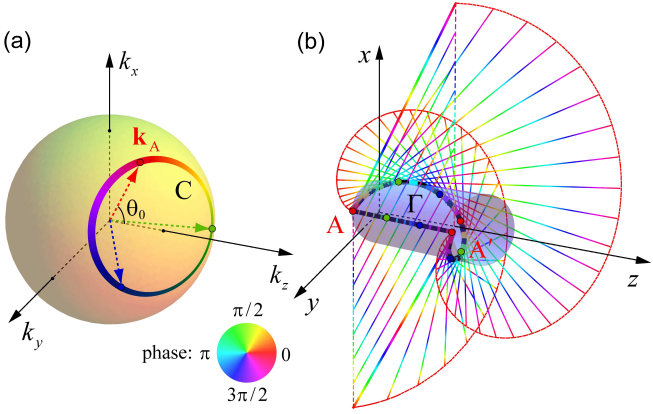}
\caption{(Color online) (a) Bessel-beam distribution (\ref{eqn:11}) on the sphere in ${\bf k}$-space with the azimuthal phase $2\pi\ell$. (b) Cylindrical caustic in the real space, an example of the closed orbit $\Gamma$ on it, and the corresponding GO rays tangent to the caustic. Scalar phases are color-coded for $\ell=-1$, $\theta_0=\pi/4$. Points $A$ and $A'$ on the caustic are connected by two paths: the straight line and the Poynting-flow helix. The phase matching yields the phase difference $2\pi\ell$ between the paths and quantization of the caustic radius. For circularly-polarized waves, the helical path brings about an additional Berry phase $\sigma\Phi_B$, Eq.~(\ref{eqn:45}).} \label{fig1}
\end{figure}

\section{Application to Bessel beams}
Importantly, our theory 
has a number of directly observable consequences. As the simplest example we take non-paraxial vector Bessel-beam solutions which are eigenmodes of $\hat J_z$ constructed from plane waves with well-defined helicity $\sigma$ (cf. \cite{EN,Li,Hawton,Bessel}). The angular spectrum of such beams is 
\begin{equation}\label{eqn:41}
{\bf \tilde E}_\ell ^\sigma  
= {\bf e}^\sigma \left( {\theta ,\phi } \right) \tilde E_\ell ^\sigma  \left( {\theta ,\phi } \right),~{\tilde E}_\ell ^\sigma  = A^\sigma  \delta \left( {\theta  - \theta _0 } \right)e^{i\ell \phi },
\end{equation}
where $A^\sigma$ is a constant amplitude, $\theta_0$ is the polar angle of conical distribution of the ${\bf k}$-vectors, Fig.~1(a), and no summation over $\sigma$ is implied here. 

For the $z$-components of OAM and SAM, Eqs.~(\ref{eqn:10}) and (\ref{eqn:11}), or (\ref{eqn:38}) and (\ref{eqn:39}), of a superposition of $\sigma =\pm 1$ beams (\ref{eqn:41}) we obtain \cite{R1}:
\begin{equation}\label{eqn:42}
{L_z } =  \ell  + \bar \sigma \frac{\Phi _B }{2\pi },~
{S_z } = \bar \sigma \left( {1 - \frac{\Phi _B }{2\pi }} \right),~
{J_z } = \ell  + \bar \sigma.
\end{equation}
Here $\bar \sigma  = ( {\left| {A^ +  } \right|^2  - \left| {A^ -  } \right|^2 } )/( {\left| {A^ +  } \right|^2  + \left| {A^ -  } \right|^2 } )$ is the averaged helicity and 
\begin{equation}\label{eqn:43}
\Phi _B  = \oint\limits_{\rm C} {{\bf A}_B  \cdot d{\bf k}}  = 2\pi \left( {1 - \cos \theta _0 } \right)
\end{equation}
is the Berry phase associated with the contour ${\rm C}=\{\theta=\theta_0, \phi\in(0,2\pi)\}$ formed by the ${\bf k}$-vectors distribution on the sphere of directions, Fig.~1(a) \cite{GP}. The Berry phase is equal to the flux of the monopole field $\mathbf{F}_B  = \partial _\mathbf{k}  \times \mathbf{A}_B  = \mathbf{k}/k^3$ through the area of the ${\bf k}$-space sphere bounded by the contour ${\rm C}$. In this manner, the $\bar\sigma$-dependent term in $L_z$ represents a monopole-flux contribution to the OAM, cf. Eq.~(87) in \cite{SUSY}. In the paraxial limit the Berry-phase terms vanish as $\Phi _B  \simeq \pi \theta _0^2  \to 0$. The values (\ref{eqn:42}) evidence an apparent partial conversion from SAM to OAM in non-paraxial light with the total AM being constant \cite{BA,Li}, akin to the spin-to-orbit AM conversion upon focusing of polarized light \cite{Li,Beksh,AMC1,AMC2,Oscar}. Indeed, in the Richards-Wolf approximation \cite{RW}, the focusing represents a geometric conical redirection of partial plane waves with their helicity being conserved. It is described exactly by the same transformation operator $\hat U (\theta,\phi)$ that describes transition to the helicity basis \cite{Oscar}.

\begin{figure}[t]
\includegraphics[width=8.6cm, keepaspectratio]{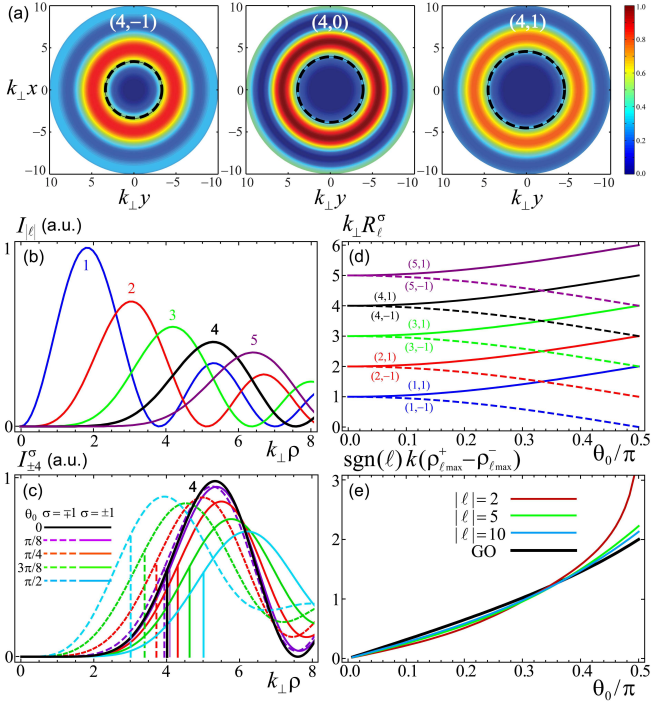}
\caption{(Color online) (a) Intensity distributions, Eq.~(\ref{eqn:45}), marked by quantum numbers $(\ell,\sigma)$ for Bessel beams with $\theta_0=3\pi /8$. The spin-dependent profiles are shown for $\ell=4$ with $\sigma=-1$, 0 (scalar case), and 1. Dashed circles indicate the GO caustics (\ref{eqn:45}). (b) Radial intensity profiles of the scalar ($\sigma=0$) or paraxial ($\theta_0\rightarrow0$) Bessel beams $I_{|\ell|}=J_{\ell}^2(\xi)$. (c) The SOI splitting of the profile of the polarized non-paraxial Bessel beam $I_{\pm4}^{\sigma}(\xi)$, Eq.~(\ref{eqn:44}), at different values of $\theta_0$; vertical lines indicate GO caustics (\ref{eqn:45}), cf. (a). (d) The GO caustics (\ref{eqn:45}) marked by ${\rm sgn}(\ell)(\ell,\sigma)$ as functions of $\theta_0$. (e) The  SOI splitting of the maxima of intensity (\ref{eqn:44}) [cf. (c)] as dependent on $\theta_0$, approaching the GO limit (\ref{eqn:45}) at $|\ell| \gg 1$.}
\label{fig2}
\end{figure}

Simultaneously with a $\sigma$-dependent OAM, the non-paraxial fields exhibit $\sigma$-dependent intensity distributions related to the modified position operator. The real-space field of the circularly-polarized Bessel beam, calculated via the Fourier transformation (\ref{eqn:A6}) of Eq.~(\ref{eqn:41}), is
\[
\mathbf{E}_\ell ^\sigma   \propto A^\sigma  \left( {\begin{array}{*{20}c}
   {\frac{{1 + \sigma }}
{2}J_\ell  \left( \xi  \right) - \sigma b\,e^{i\left( {\sigma  - 1} \right)\varphi } J_{\ell  + \sigma  - 1} \left( \xi  \right)}  \\
   {\frac{{1 - \sigma }}
{2}J_\ell  \left( \xi  \right) + \sigma b\,e^{i\left( {\sigma  + 1} \right)\varphi } J_{\ell  + \sigma  + 1} \left( \xi  \right)}  \\
   { - i\sigma \sqrt {2ab} e^{i\sigma \varphi } J_{\ell  + \sigma } \left( \xi  \right)}  \\
 \end{array} } \right)e^{ik_{\parallel}  z + i\ell \varphi },
\]
where $\left( {\rho ,\varphi ,z} \right)$ are the cylindrical coordinates in real space, $a = \cos ^2 \left( {\theta _0 /2} \right)$, $b = \sin ^2 \left( {\theta _0 /2} \right)$, $\xi=k_\bot\rho$, $k_ \bot = k\sin \theta _0$, $k_{\parallel} = k\cos \theta _0$, $J_n (\xi)$ are the Bessel functions of the first kind, and the field components are written in the basis $\left(\frac{{\bf e}_x+i{\bf e}_y}{\sqrt{2}},\frac{{\bf e}_x-i{\bf e}_y}{\sqrt{2}},{\bf e}_z\right)$. This field has a cylindrically symmetric intensity distribution, $I_\ell ^\sigma   = \left| {\mathbf{E}_\ell ^\sigma  } \right|^2$, given by
\begin{equation}\label{eqn:44}
I_\ell ^\sigma  
\propto \left| {A^\sigma  } \right|^2 \left[ {a^2 J_\ell^2 \left( \xi \right) + b^2 J_{\ell + 2\sigma }^2 \left( \xi \right) + 2abJ_{\ell + \sigma }^2 \left( \xi \right)} \right],
\end{equation}
Below we show that the polarization-dependent intensity distributions (\ref{eqn:44}) (see Fig.~2(a)) signify the SOI of light.

The $\ell$- and $\sigma$-dependence of the radial intensity profile (\ref{eqn:44}) can be explained via a geometrical-optics (GO) ray picture and the \textit{quantization of caustic} underlying the maximum of the intensity. The rays associated with a Bessel beam are those that form an angle $\theta_0$ with the $z$-axis and touch a cylindrical caustic of radius $\rho  = R_\ell ^\sigma$  \cite{Berry2008}, Fig.~1. The quantization condition for a closed orbit $\Gamma$ is $\oint\limits_\Gamma  {\bf k}  \cdot d{\bf r} = 2\pi \ell$. Using the underlying position (\ref{eqn:7}), ${\bf r'}^\sigma   = {\bf r} - \sigma {\bf A}_B$, we observe that the Berry phase changes the effective optical length of a closed orbit on the cylindrical surface, Fig~1(b). For the orbit $\Gamma  = \left\{ {\rho  = R_\ell ^\sigma  ,\varphi  \in \left( {0,2\pi } \right)} \right\}$ it becomes $k_ \bot  \left[ {2\pi \, {\rm sgn}(\ell) R_\ell ^\sigma   - \sigma {\kern 1pt} \Phi _B } \right]$, which yields
\begin{equation}\label{eqn:45}
k_ \bot  R_\ell ^\sigma   = \left| \ell  + \sigma {\kern 1pt} \frac{\Phi _B }{2\pi } \right|~.
\end{equation}
Similar Berry-phase effects appear in quantum quantization problems \cite{Quantization}, e.g., the half-integer Hall effect in graphene \cite{Graphene}. Note also the exact correspondence between the GO caustic (\ref{eqn:45}) and the wave OAM (\ref{eqn:42}), $|L_z|=k_{\bot}R_\ell ^\sigma$, which reflects the OAM interpretation as ${\bf r}\times{\bf k}$ for the rays. Figure 2 shows $\ell$- and $\sigma$-dependent intensity distributions (\ref{eqn:44}) of the Bessel beams vs. the GO caustics (\ref{eqn:45}). Spin-dependent splitting of caustics and intensity maxima are the optical analogues of the fine spin-orbit splitting of levels in quantum systems. The $\sigma$-dependence in radial distributions of non-paraxial vortex fields can be observed experimentally by tightly focusing paraxial light with different polarizations, cf. \cite{Gorod}.

\begin{figure}[t]
\includegraphics[width=8.6cm, keepaspectratio]{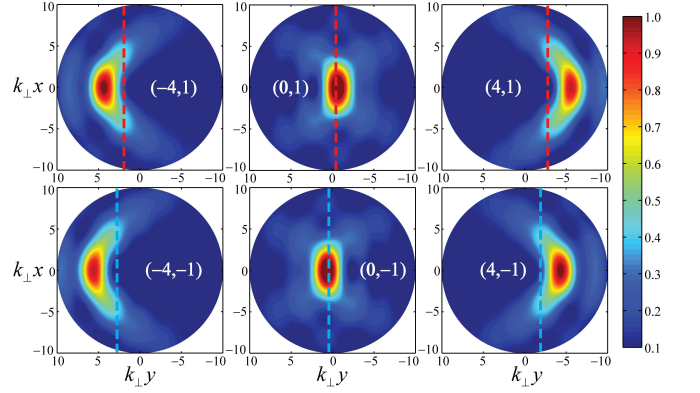}
\caption{(Color online) Transverse intensity distributions at $z=0$ of the asymmetric Bessel beams with $\delta=\pi/2$ and $\theta_0 =3\pi/8$ marked by quantum numbers $(\ell,\sigma)$. Dashed lines indicate the $\ell$- and $\sigma$-dependent transverse shifts of the centers of gravity, Eq.~(\ref{eqn:46}), i.e., orbital and spin Hall effects of light in free space. We have verified that the centers of gravity calculated numerically from the intensity distributions and theoretically from Eq.~(\ref{eqn:46}) coincide.} \label{fig3}
\end{figure}

Finally, we consider the Hall effects of light, which are described by the position (\ref{eqn:12}). For this purpose we break the symmetry of the Bessel beams (\ref{eqn:41}) along the $k_x$-axis and assume that the plane-wave components are distributed in the range $\phi\in (-\delta,\delta)$, $0<\delta <\pi$. (Such truncated azimuthal distributions can be generated via focusing by the corresponding sector of a lens \cite{Half}.) Substitution of this distribution in Eqs.~(\ref{eqn:12}) and (\ref{eqn:13}), or (\ref{eqn:36}) and (\ref{eqn:A9}), shows mutually orthogonal tilt and displacement of the beam:
\begin{equation}\label{eqn:46}
P_x =\gamma k_{\bot}~,~~k_\bot Y_{\ell}^{\sigma}= - \gamma \left( \ell  + \bar \sigma \frac{{\Phi _B }}{{2\pi }}\right)~.
\end{equation}
Here $\gamma = (\sin\delta)/\delta$, $X(z)=z P_x/P_z$ ($P_z=k_\parallel$), and the second expression (\ref{eqn:46}) closely resembles Eqs.~(\ref{eqn:42}) and (\ref{eqn:45}). The $\ell$- and $\sigma$-dependent parts of the transverse shift of the center of gravity of the beam, $Y_{\ell}^{\sigma}$, describe the orbital and spin Hall effects of light in free space, Fig.~3. A related spin-Hall effect has been observed upon focusing of light with a ``half-lens'' ($\ell=0$ for $\delta=\pi/2$) \cite{Half}, whereas the orbital-Hall effect can be measured in a similar manner by focusing  vortex beams with broken symmetry. The values of $L_z$ and $S_z$ for the asymmetric beam are given by the same Eq.~(\ref{eqn:42}), but in this case the OAM has an extrinsic contribution , $L_z^{\rm ext}=-P_x Y_{\ell}^{\sigma}$ (\ref{eqn:14}): 
\begin{equation}\label{eqn:47}
L_z^{\rm ext}=\gamma^2 L_z,~L_z^{\rm int}=(1-\gamma^2) L_z.
\end{equation}
Hence, the Hall effects of light can be interpreted as an \textit{intrinsic-to-extrinsic OAM conversion} \cite{SHE1,OHE1} which is also accompanied by generation of a transverse OAM component $L_x^{\rm ext}=P_z Y_{\ell}^{\sigma}=- \gamma\cot\theta_0 L_z$ \cite{Aiello}. The total conversion is achieved at $\delta\rightarrow 0$, $\gamma\rightarrow 1$.

\section{Conclusion}
To summarize, we have revisited the problem of the identification of the spin and orbital angular momenta of nonparaxial light in free space. It has been shown that this issue is closely related to the determination of the position of the center of gravity of a light beam or a wave packet. We have given an exact self-consistent solution to these problems in terms of quantum-operator formalism and using classical Poynting energy flows. In the helicity representation, taking into account the transverse nature of the electromagnetic fields, the operators of the OAM, SAM, and position become diagonal, but exhibit non-canonical commutation relations. We have shown that the unusual features of these operators originate from the Berry-phase terms and can be associated with manifestations of the spin-orbit interaction of light. Indeed, anomalous Berry terms in the OAM and position operators describe spin-dependent part of OAM (responsible for spin-to-orbital AM conversion) and spin-dependent shift of the center of gravity of light (i.e., the spin-Hall effect of light). We have applied the general theory to symmetric and asymmetric vector Bessel beams and found that our non-canonical operators indeed correspond to the observable quantities. The obtained Bessel-beam intensity distributions exhibit fine SOI splitting of caustics and Hall effects of light in perfect agreement with the derived OAM and position operators. These effects can be observed experimentally in tightly focused fields.

This work was supported by the European Commission (Marie Curie Action), Science Foundation Ireland (Grant No. 07/IN.1/I906), the Australian Research Council (ARC), and von Humboldt foundation. We are grateful to A. Y. Bekshaev and M. V. Berry for fruitful discussions.

\renewcommand{\theequation}%
{A\arabic{equation}}
\renewcommand{\thefigure}%
{A\arabic{figure}}
\setcounter{equation}{0}
\setcounter{figure}{0}

\appendix
\section{Operator formalism for wave packets and beams}

One can separate two basic situations, for which the operator formalism of Section II can be adopted in a slightly different way. The first one is evolution of a wave-packet-like field localized in 3D space. Obviously, such field is nonmonochromatic and time-dependent. The plane-wave Fourier decomposition of the complex electric field can be written as
\begin{equation}
\label{eqn:A1}
{\bf E}\left( {\bf r},t \right) = \frac{g}{(2\pi)^{3/2} }\int {{\bf \tilde E}\left( {\bf k} \right)e^{i{\bf k} \cdot {\bf r}-i\omega({\bf k})t} } d^3 {\bf k}~,
\end{equation}
where $d^3{\bf k}=dk_x dk_y dk_z = k^2\sin\theta dk d\theta d\phi$, $\omega({\bf k})=k$ is the dispersion relation, factor $g=\sqrt{2\omega/\varepsilon_0}$ ($\varepsilon_0$ is the vacuum permittivity) is introduced for proper normalization of energy below, and the real wave electric field is given by ${\bm {\mathcal E}}\left( {\bf r},t \right) = {\rm Re}\, {\bf E}\left( {\bf r},t \right)$. In the helicity basis one has
\begin{equation}
\label{eqn:A2}
{\bf \tilde E}\left( {\bf k} \right) = \tilde E^{+} \left( {\bf k} \right){\bf e}^{+} \left( {\bf k} \right) +
\tilde E^{-} \left( {\bf k} \right){\bf e}^{-} \left( {\bf k} \right)~.
\end{equation}
The energy of the wave-packet field is given by the 3D space integral of the intensity (we omit inessential constant factors) and can be written as:
\begin{eqnarray}\label{eqn:A3}
\nonumber
W = \frac{1}{2}\int {\left(\varepsilon_0{\left| \bm{\mathcal E} \right|}^2+ \mu_0{\left| \bm{\mathcal H} \right|}^2\right)} {d^3}{\mathbf{r}} \\
\nonumber
= \frac{1}{4}\int {\left(\varepsilon_0{{\left| \bf{E} \right|}^2}+ \mu_0{{\left| \bf{H} \right|}^2}\right)} {d^3}{\mathbf{r}} \\
= \int {\omega{{\tilde E}^{\sigma *}}{{\tilde E}^\sigma }{d^3}{\mathbf{k}}}  \equiv \left\langle {{{\tilde E}^\sigma }} \right|\omega \left| {{{\tilde E}^\sigma }} \right\rangle.
\end{eqnarray}
Here ${\bm {\mathcal H}}\left( {\bf r},t \right) = {\rm Re}\, {\bf H}\left( {\bf r},t \right)$ is the magnetic field, $\mu_0$ is the vacuum permeability, $d^3{\bf r} = dx dy dz$, summation over $\sigma=\pm 1$ is implied hereinafter, and we performed some standard calculations with Maxwell equations and the Fourier transform (\ref{eqn:A1}). Thus, the convolution implies 3D integration of the field spectral amplitudes over the ${\bf k}$-space. At the same time, to determine properly the state vector $\left| {\tilde E^\sigma  } \right\rangle$, one has to take into account the temporal dependence of the field, namely: 
\begin{equation}\label{eqn:A4}
\left| {\tilde E^\sigma  } \right\rangle  = \tilde E^\sigma  \left( \mathbf{k} \right)e^{-i\omega \left( \mathbf{k} \right)t}~.
\end{equation}
We assume normalization which has the meaning of the unit number of photons in the wave packet: $N= \left\langle {{{\tilde E}^\sigma }} \right| \left. {{{\tilde E}^\sigma }} \right\rangle =1$. Substituting the state vector (\ref{eqn:A4}) with the definition of convolution (\ref{eqn:A3}) into Eq.~(\ref{eqn:12}), we obtain the time-dependent position of the center of gravity of the wave packet moving in space:
\begin{equation}\label{eqn:A5}
\mathbf{R}\left( t \right) =  - \operatorname{Im} \int {\tilde E^{\sigma *} \partial _\mathbf{k} \tilde E^\sigma  } d^3 \mathbf{k} - \int {\sigma \mathbf{A}_B \left| {\tilde E^\sigma  } \right|^2 } d^3 \mathbf{k} + \mathbf{V}t,
\end{equation}
where the velocity of the wave-packet motion is given by 
\[
\mathbf{V} = \int {(\partial _\mathbf{k} \omega) \left| {\tilde E^\sigma  } \right|^2 } d^3 \mathbf{k} = \int {\bm{\kappa }\left| {\tilde E^\sigma  } \right|^2 } d^3 \mathbf{k}. 
\]
Note that the same expression for the wave-packet center can be obtained by convolution of the canonical coordinate operator $\mathbf{\hat r} = i\partial _\mathbf{k}$ with the vector state $\left| {\tilde {\bf E}^\sigma  } \right\rangle  = \tilde {\bf E}^\sigma  \left( \mathbf{k} \right)e^{-i\omega \left( \mathbf{k} \right)t}$. The Berry-connection term arises in this case from the $\partial _\mathbf{k}$ derivatives of the helicity basic vectors ${\bf e}^{\sigma}$, Eq.~(\ref{eqn:9}). The linear and angular momenta of the field, Eqs.~(\ref{eqn:10}), (\ref{eqn:11}), and (\ref{eqn:13}), can be calculated in a manner similar to Eqs.~(\ref{eqn:A3})--(\ref{eqn:A5}).

The second typical problem that arises in optics deals with a beam-like monochromatic field ($\omega=k=const$) propagating in the positive $z$-direction and localized only in the transverse $(x,y)$-dimensions. In this case, it is natural to use the (2+1)D version of quantum-like formalism, where $z$ instead of time plays the role of the independent variable, whereas ${\bf r}_{\perp}=(x,y)$ is the effective 2D space allowing normalization of the transverse field distributions \cite{Aiello2}. Because of the monochromaticity, only two components of the ${\bf k}$-vector are independent, and the $z$-component can be expressed as $k_z  = k_z \left( {\mathbf{k}_ \bot  } \right) = \sqrt {\omega ^2  - k_ \bot ^2 }$, ${\bf k}_\bot=(k_x,k_y)$. This determines the following 2D plane-wave Fourier decomposition of the complex time-independent electric field \cite{Aiello2}:
\begin{equation}\label{eqn:A6}
{\bf E}\left( {\bf r}_{\perp},z \right) = \frac{g}{{2\pi }}\int {{\bf \tilde E}\left( {\bf k}_{\perp} \right)e^{i{\bf k}_{\perp} \cdot {\bf r}_{\perp} +ik_z({\bf k}_{\perp})z} } d^2 {\bf k}_{\perp}~,
\end{equation}
where the real wave electric field is given by ${\bm{\mathcal E}}\left( {\bf r},t \right) = {\rm Re} \left[ {\bf E}\left( {\bf r}\right)e^{-i\omega t} \right]$ and the element of the 2D area of integration is ${d^2}{\mathbf{k}}_{\perp} = d{k_x}d{k_y}$. Alternatively, one can use $\mathbf{\tilde E} = \mathbf{\tilde E}\left( {\theta ,\phi } \right)$ and ${d^2}{\mathbf{k}}_{\perp} = {k^2}\cos \theta \sin \theta d\theta d\phi$ in spherical coordinates with two independent dimensions $(\theta,\phi)$. The characteristic energy of the wave beam is, in fact, the energy per unit $z$-length which is obtained by the 2D integration of the time-averaged intesity over $d^2{\bf r}_{\bot} = dx dy$:
\begin{eqnarray}\label{eqn:A7}
\nonumber
W = \frac{1}{2}\int {\left(\varepsilon_0\overline{{\left| \bm{\mathcal E} \right|}^2}+ \mu_0\overline{{\left| \bm{\mathcal H} \right|}^2}\right)} {d^2}{\mathbf{r}}_{\bot} \\
\nonumber
= \frac{1}{4}\int {\left(\varepsilon_0{{\left| \bf{E} \right|}^2}+ \mu_0{{\left| \bf{H} \right|}^2}\right)} {d^2}{\mathbf{r}}_{\bot} \\
= \int {\omega{{\tilde E}^{\sigma *}}{{\tilde E}^\sigma }{d^2}{\mathbf{k}}_{\bot}}  \equiv \left\langle {{{\tilde E}^\sigma }} \right|\omega \left| {{{\tilde E}^\sigma }} \right\rangle=\omega~.
\end{eqnarray}
Here ${\bm{\mathcal H}}\left( {\bf r},t \right) = {\rm Re} \left[ {\bf H}\left( {\bf r}\right)e^{-i\omega t} \right]$, the overline stands for the time averaging, and we assumed the unit number of photons per unit $z$-length in the beam: $N= \left\langle {{{\tilde E}^\sigma }} \right| \left. {{{\tilde E}^\sigma }} \right\rangle =1$.
Thus, the convolution for beam-like fields implies 2D integration over $(k_x,k_y)$ or $(\theta,\phi)$ in the ${\bf k}$-space (these are equivalent unless we consider evanescent modes). To determine properly the state vector $\left| {\tilde E^\sigma  } \right\rangle$, one has to take into account the $z$-dependence of the field, cf. Eq.~(\ref{eqn:A4}): 
\begin{equation}\label{eqn:A8}
\left| {\tilde E^\sigma  } \right\rangle  = \tilde E^\sigma  \left( \mathbf{k}_{\bot} \right)e^{ik_z \left( \mathbf{k}_{\bot} \right)z}~.
\end{equation}
Substituting definitions (\ref{eqn:A7}) and (\ref{eqn:A8}) into Eq.~(\ref{eqn:12}), we obtain the $z$-dependent transverse position of the center of gravity of the propagating wave beam \cite{Aiello2}:
%
\begin{eqnarray}\label{eqn:A9}
\nonumber
\mathbf{R}_ \bot  \left( z \right) =  - \operatorname{Im} \int {\tilde E^{\sigma *} \partial _{\mathbf{k}_ \bot  } \tilde E^\sigma  } d^2 \mathbf{k}_ \bot \\
- \int {\sigma \mathbf{A}_B \left| {\tilde E^\sigma  } \right|^2 } d^2 \mathbf{k}_ \bot  
+ \mathbf{V}z~,
\end{eqnarray}
%
where the `velocity' of the motion along $z$ is given by 
\[
\mathbf{V} =  - \int { (\partial _{\mathbf{k}_ \bot  } k_z) \left| {\tilde E^\sigma  } \right|^2 } d^2 \mathbf{k}_ \bot   = \int {\frac{\bm{\kappa }}
{{k_z }}\left| {\tilde E^\sigma  } \right|^2 } d^2 \mathbf{k}_ \bot. 
\]
The linear and angular momenta (more precisely, their values per unit $z$-length) are calculated from Eqs.~(\ref{eqn:10}), (\ref{eqn:11}) and (\ref{eqn:13}) in a similar manner (see also Sections III and IV). Note that despite the above $(2+1)D$ quantum-like formalism, they are vectors in 3D space. This does not cause any difficulties if one uses $\hat{z}=i\partial_{k_z} = z$ which yields $Z=z$ in the 3D calculations, cf. \cite{Aiello}.

It should be emphasized that, despite our using the same letters for the unifying formalism, quantities ${\bf E}$, $\tilde{\bf E}$, $W$, ${\bf R}$, etc. have different meanings for the 3D-localized wave-packet polychromatic fields, Eqs.~(\ref{eqn:A1})--(\ref{eqn:A5}), and 2D-localized monochromatic beams, Eqs.~(\ref{eqn:A6})--(\ref{eqn:A9}).


\begin{thebibliography}{99}
\bibitem{R1} As it is known, the Bessel beams cannot be normalized and we calculate the finite ratio of the quantities Eqs.~(\ref{eqn:10})--(\ref{eqn:14}) to the infinite norm $N=\left\langle {\tilde E^\sigma  } \right|\left. {\tilde E^\sigma  } \right\rangle$.
\end{thebibliography}
\end{document}